\documentclass{PoS}

\title{The effect of radial gas flows on the chemical evolution of the Milky Way and M31}

\ShortTitle{The Milky Way and M31}

\author{\speaker{Emanuele Spitoni}\\
        Dipartimento di Fisica,  Universit\`a di Trieste\\
        E-mail: \email{spitoni@oats.inaf.it}}

\author{Francesca Matteucci\\
         Dipartimento di Fisica,  Universit\`a di Trieste\\
        E-mail: \email{matteucci@oats.inaf.it}}

\abstract{We present detailed chemical evolution models
  for the Milky Way and M31 in presence of radial gas flows. These
  models follow in detail the evolution of several chemical elements
  (H, He, CNO, $\alpha$ elements, Fe-peak elements) in space and time. The
  contribution of supernovae of different type to chemical enrichment
  is taken into account. We find
  that an inside-out formation of the disks coupled with radial gas
  inflows of variable speed can reproduce very well the observed
  abundance gradients in both galaxies. We also discuss the effects of
  other parameters, such as a threshold in the gas density for star
  formation and efficiency of star formation varying with galactic
  radius.  Moreover, for the first time we compute the galactic
  habitable zone in our Galaxy and M31 in presence of radial gas
  flows. The main effect is to enhance the number of stars hosting a
  habitable planet with respect to the models without radial flow, in
  the region of maximum probability for this occurrence. In the Milky
  Way the maximum number of stars hosting habitable planets is at 8
  kpc from the Galactic center, and the model with radial gas flows
  predicts a number of planets  which is 38\% larger than that predicted by the
  classical model.}

\FullConference{XIII Nuclei in the Cosmos,\\
		7-11 July, 2014\\
		Debrecen, Hungary }

\begin{document}

\section{Introduction}
The majority of chemical evolution models assumes that galactic
disks form by means of infall of gas and divides the disk into several
independent rings. However, if the infall is important then radial gas
flows should be taken into account as a dynamical consequence of
infall. The infalling gas has a lower angular momentum than the
circular motions in the disk, and mixing with the gas in the disk
induces a net radial inflow (Lacey \& Fall 1985). In this contribution
we will discuss the effects of radial gas flow on the chemical
evolution of the Milky and M31, and on the galactic
habitable zone of those galactic systems.

\section{The oxygen abundance gradients for the Milky Way and M31 in presence of radial gas flows}

In Fig. 1 we show our results for the abundance gradient for oxygen in
presence of radial gas flows of the Milky Way and M31,
respectively. In the left panel we show the best model of Mott at
al. (2013) for the Milky Way compared with the data from Cepheids. 
\begin{figure}
	  \centering  
     \includegraphics[scale=0.3]{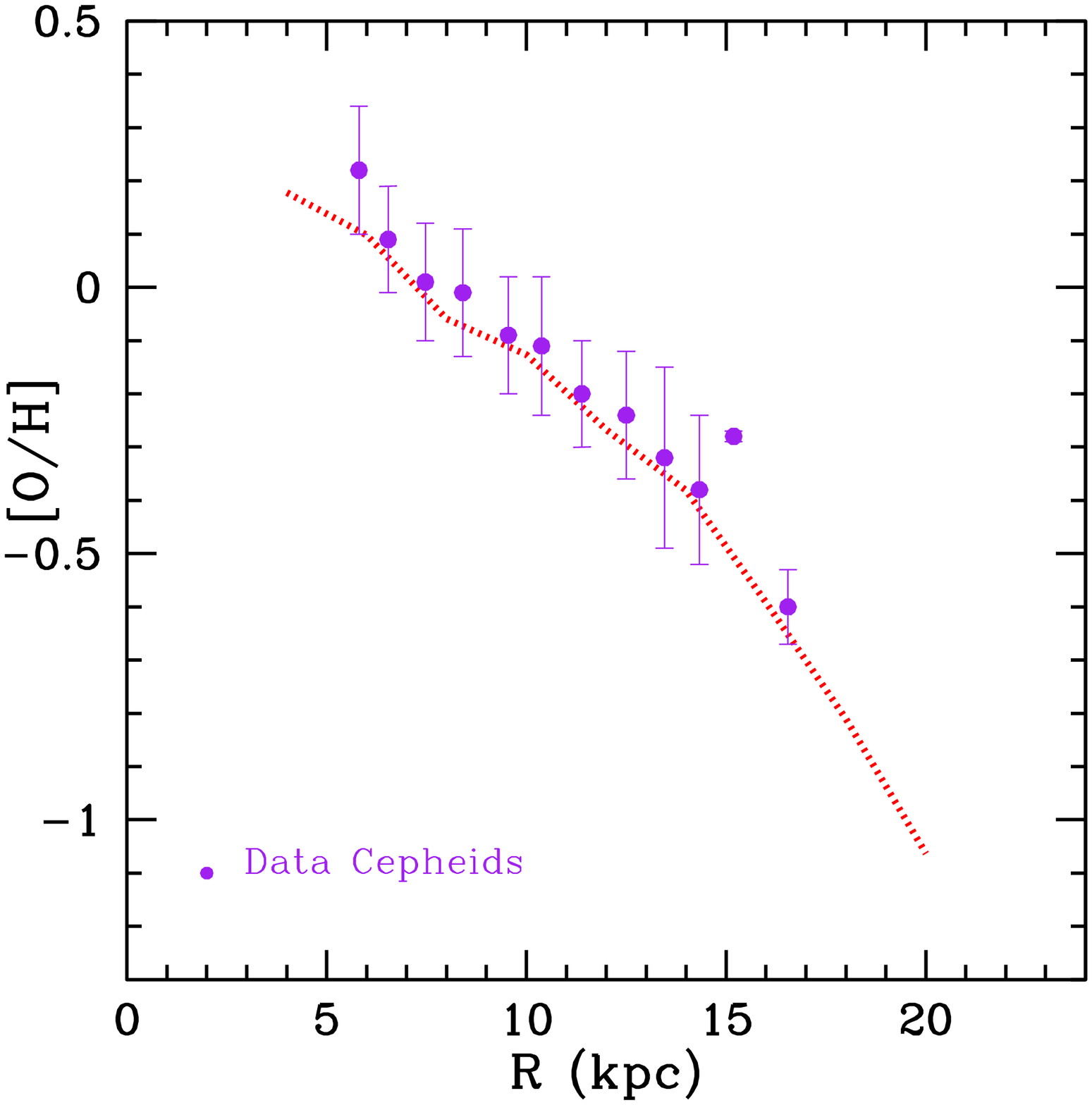} 
   \includegraphics[scale=0.56]{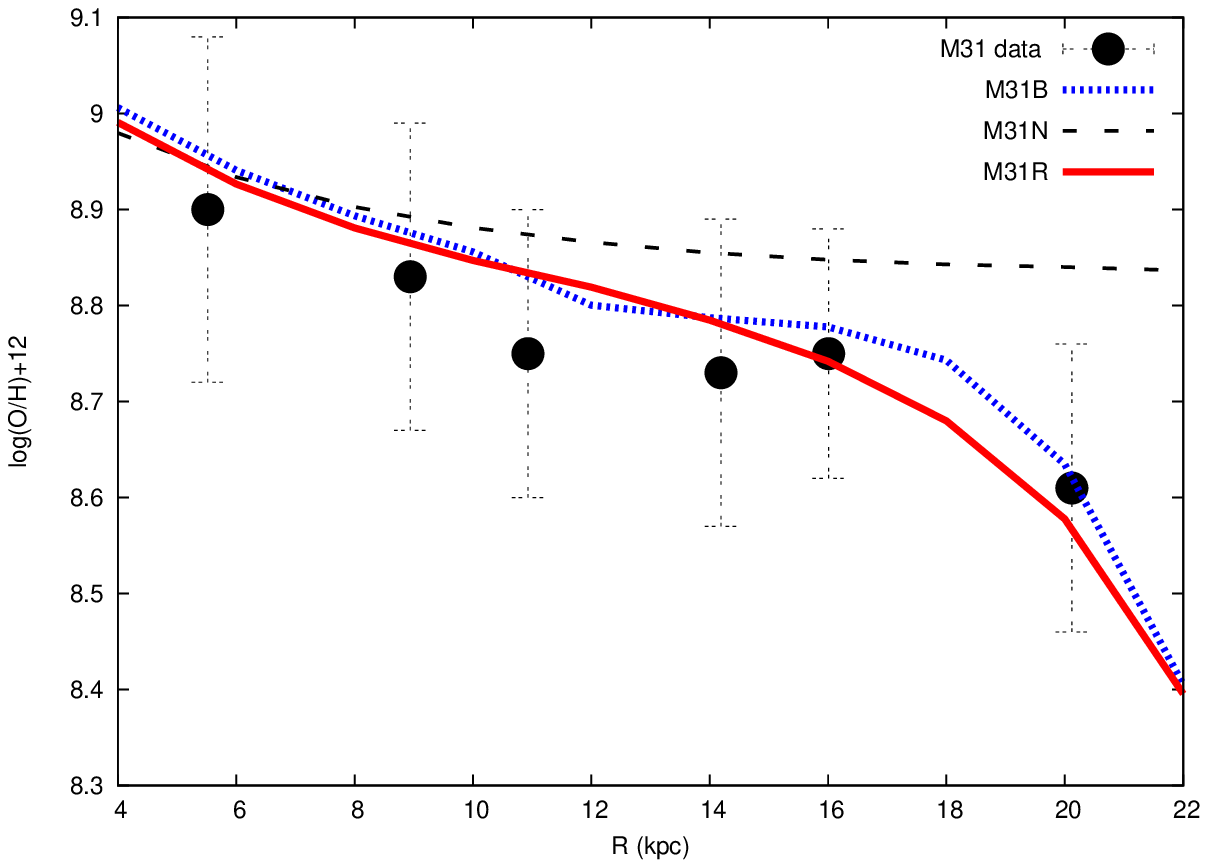} 
   
    \caption{ {\it Left panel}: Oxygen gradient in the Milky Way. The
      dotted line represents model of Mott et al. (2013) with radial
      gas flows (model MW-R in Table 1). The model is compared with
      the data from Cepheids of Luck \& Lambert (2011). {\it Right
        panel}: Oxygen gradient in M31. The dotted line represents a
      model without radial gas flows and no threshold in the gas
      (model M31-N in Table 1), whereas the dashed line represents the
      best model with gas flows and no threshold (model M31-R in Table
      1), We also report the best model M31-B for a ``static'' model
      without radial as flow but in presence of threshold.  The
      results are compared with the data from HII regions and
      supernova remnants (see Spitoni et al. 2013 for references). }
		\label{aD}
\end{figure} 
For M31 the best model of Spitoni et al. (2013) is compared with data
from supernova remnants, and HII regions (see Spitoni et al.  for
references). For both models we found that an inside-out formation of
the disk, with no threshold in the surface gas density for the star
formation rate (SFR) coupled with radial gas inflows of variable speed
(in the left panel of Fig. 2 we show the radial gas inflow velocities
for the Milky Way and M31) can reproduce very well the observed
abundance gradients. 

In Table 1 we report the properties of the best
models for Milky Way model (MW-R) and the one for M31 (M31-R) in
presence of radial flows. In the same  Table  all the models
considered in this work are reported.  If a threshold in the gas
density  is assumed its values is shown in  the second column, expressed
in $M_{\odot} pc^{-2}$. In column 3, the inside-out scenario is
expressed by a linear variation of the time-scale of infall
$\tau_D(R)$, and if this assumption is absent, the time-scale is set
to be constant.  In column 4 the different prescriptions for the SFE
$\nu$ are reported. In the last column the presence of radial gas
flows is indicated.
\begin{table*}

\caption{The list of the models described in this work.}
\scriptsize

\label{models}
\begin{center}
\begin{tabular}{c|ccccc}
  \hline
\hline
\\
 Model &Threshold [M$_{\odot}$ pc$^{-2}$]& $\tau_d$ (I-O) [Gyr] &SFE $\nu$ [Gyr$^{-1}$]&Radial inflow  \\
\\
\hline

M31-R& / &  0.62 R (kpc) +1.62 Gyr &2&velocity pattern Fig. 2\\

\hline
M31-N& / &  0.62 R (kpc) +1.62 Gyr &2&/\\

\hline

M31-B& 5 &  0.62 R (kpc) +1.62 Gyr &2&/\\

\hline
MW-R& / &  1.033 R (kpc) -1.27 Gyr &1&velocity pattern Fig. 2\\

\hline

MW-A& 7 (Thin Disk) &  1.033 R (kpc) -1.27 Gyr &1&/\\
& 4 (Halo-Thick Disk)&&&\\
\hline
MW-B& 7 (Thin Disk) &  1.033 R (kpc) -1.27 Gyr &$\nu(R) \propto R^{-1}$&/\\
& 4 (Halo-Thick Disk)&& &\\
\hline
MW-C& / &  1.033 R (kpc) -1.27 Gyr &1&/\\
\hline

MW-D& 7 (Thin Disk) &  3 Gyr &1&/\\
& 4 (Halo-Thick Disk)&&&\\
\hline

\end{tabular}
\end{center}

\end{table*}

  For the Milky Way we also discuss the formation and the temporal
   evolution of the abundance gradient for the oxygen. Recently,
   Cresci et al. (2010) measured oxygen abundances across three
   star-forming galaxies at redshift z = 3. The most striking result
   of this study is the existence of a positive gradient in the oxygen
   abundance. In other words the O abundance in the inner disks of
   these galaxies seems to decrease towards the galactic center.  In
   the right panel of Fig, 2 we show the temporal evolution for our
   best model in presence of radial gas inflows (model MW-R). In accordance with
   results of Cresci et al. (2010) (z=3 corresponding to a cosmic time
   of 2 Gyr from the Big Bang) our model shows an increase of
   metallicity from the outer regions up to 8 kpc where it reaches a
   peak and then a decrease for R < 8 kpc towards the Galactic centre.
   Our explanation for the gradient inversion in the Milky Way is
   based on the inside-out disc formation: (i) at early epoch (z = 3)
   the efficiency of chemical enrichment (i.e. of the SFR) in the
   inner regions is high but the rate of infalling primordial gas is
   dominating, thus diluting the gas more in the inner than in the
   outer regions; (ii) as time passes by, the infall of pristine gas
   in the inner parts decreases and the chemical enrichment takes
   over; (iii) then, at later epochs, the SFR in the inner regions is
   still much higher than in the outer parts of the disc where the gas
   density is very low, but the infall is lower and the abundance
   gradients become negative also in the inner regions.

\section{The effects of several parameters on the abundance gradient of the Milky Way in absence of radial flows}
In Fig. 3 we show our results for the present day oxygen gradient in
the Milky Way showing the effects of different parameters on models
without radial gas flows, compared with the data from Cepheids.  We
examine the following parameters: inside-out formation, threshold in
gas density for the SFR, and the star formation efficiency (SFE).  The
model  with inside-out formation, threshold and constant SFE (model MW-A) well reproduce  the abundance gradient up to 14 kpc.
\begin{figure}
	  \centering   
\includegraphics[scale=0.3]{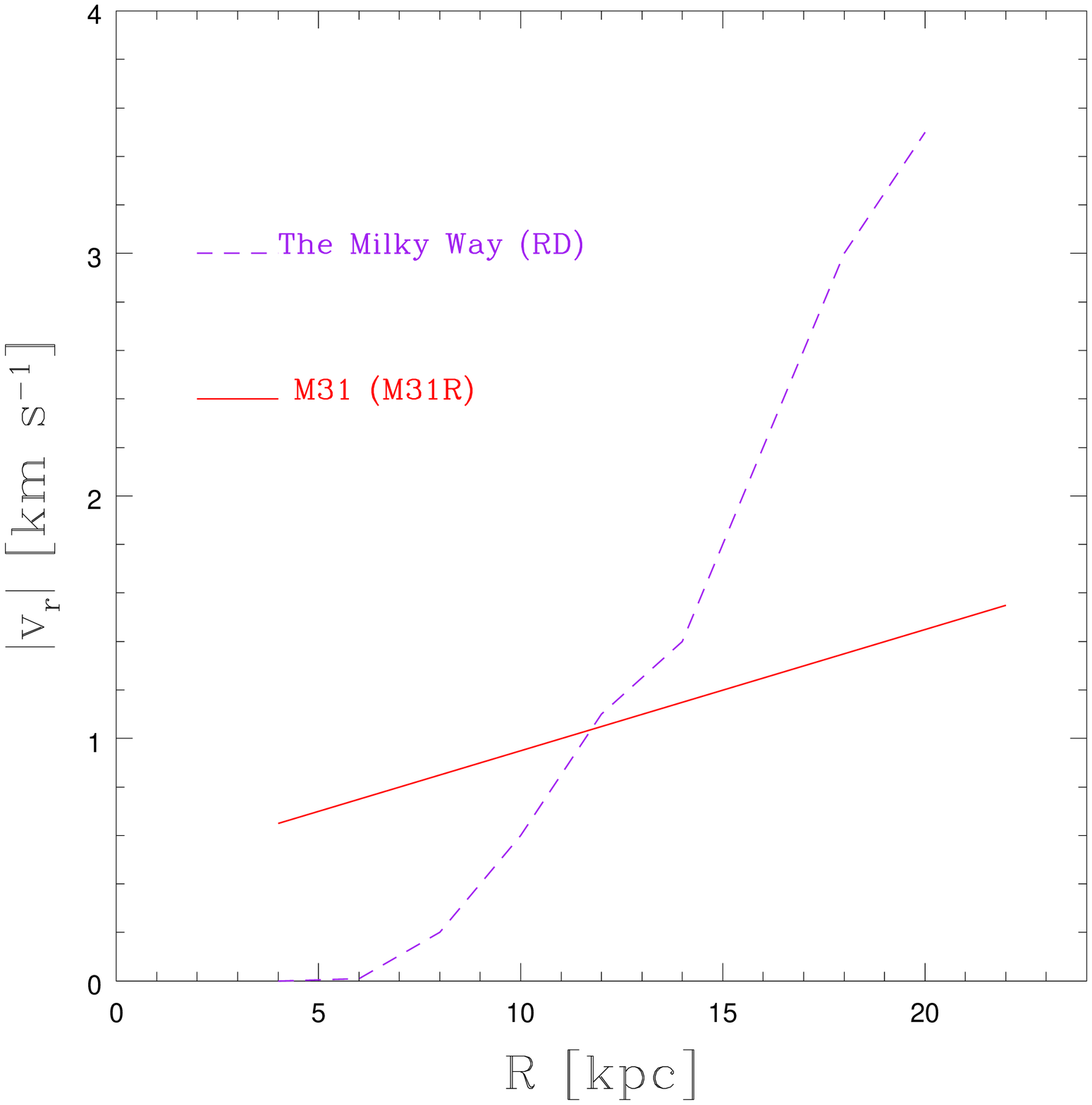} 
    \includegraphics[scale=0.3]{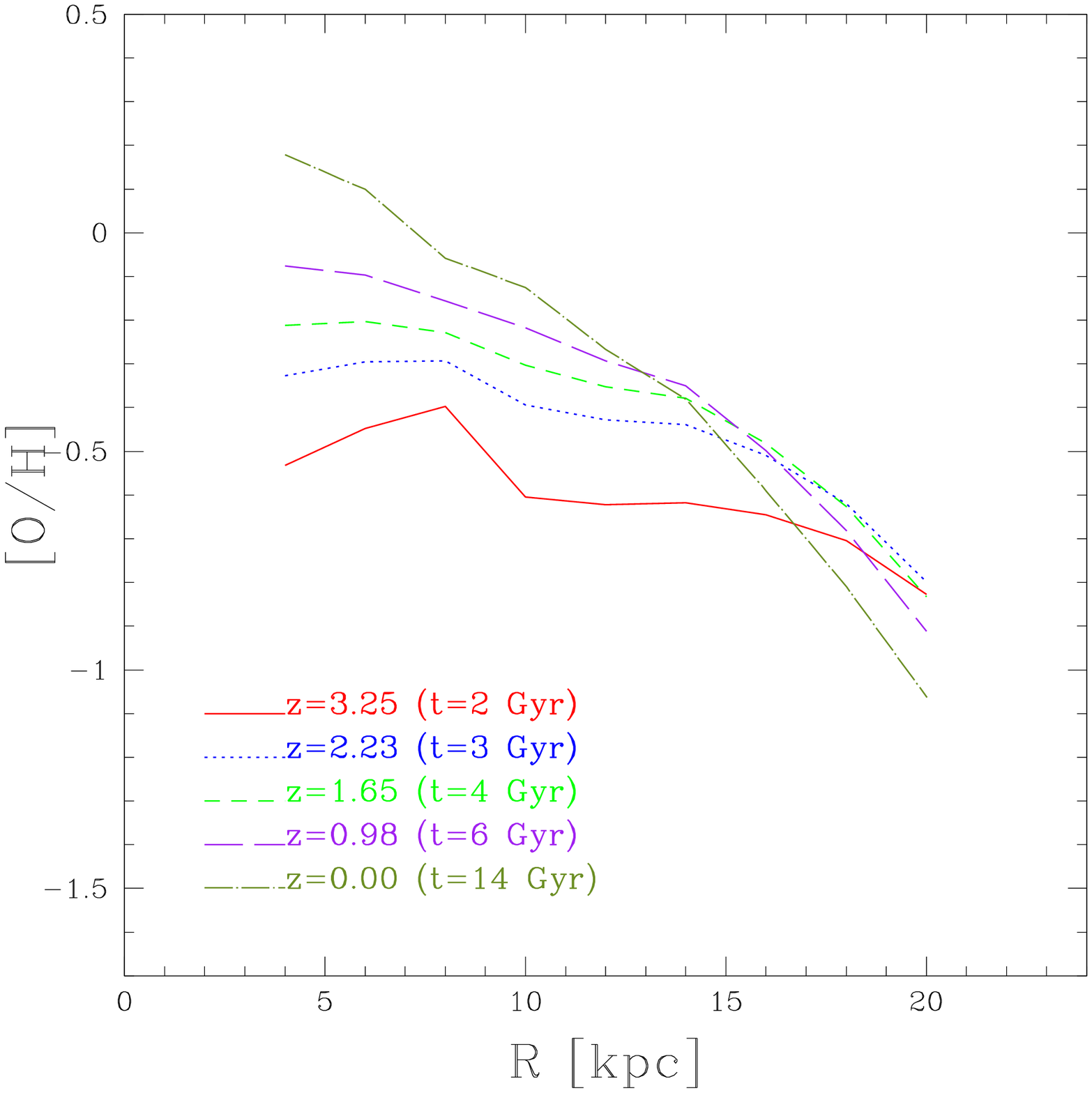} 
 
    \caption{{\it Left panel}: velocity pattern for the radial inflows
      of gas for the Milky Way model MW-R (dashed line) and for the M31
      model one  M31-R (solid line) (The properties of those models are reported in Table 1).  {\it Right panel}: Evolution with
      redshift of the abundance gradients for the best model MW-R for
      the oxygen.  The evolution is studied by computing the abundance
      gradients for the redshifts z=3.25, 2.23, 1.65, 0.98 and 0.0
      whose correspond to the times t=2, 3, 4, 6 and 14 Gyr after the
      Big Bang (in a $\Lambda CDM$ cosmology).}
		\label{altoredshiftORD}
\end{figure} 
If instead, we do not assume an inside-out formation for the thin
disk, and keep constant the timescale of infall $\tau_D$ , the present
day abundance gradients provided by the model are too flat in the
inner part of the disk even if a threshold in the gas density is
assumed (model MW-D in Table 1).  In fact, the threshold influences
mostly the outer gradient.  The model with inside out formation but no
threshold (model MW-C in Table 1) shows an abundance gradient too flat
between 6-12 kpc and in the outer parts of the disk it even increases,
clearly at variance with the observational data.  This is in agreement
with what found in Chiappini et al. (2001).  Thus, we can conclude
that a threshold in the gas density seems to be necessary to have the
right trend of the gradients in the outer parts of the disk in a model
without radial gas flows.  The model MW-B in Table 1 which assumes a
variable SFE, a threshold, and inside-out formation, provides a good
fit to the observed present day abundance gradients from 4 to 14 kpc
in the Milky Way. However, beyond this distance the gradient predicted
by the models is too flat and inconsistent with the observations.

\begin{figure}
	  \centering   
 \includegraphics[scale=0.35]{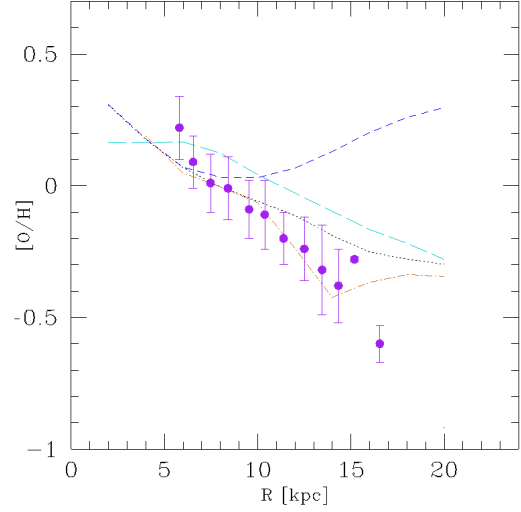} 
    \caption{Effects on the oxygen abundance gradients of several
      parameters that characterize the chemical evolution in absence
      of radial flows.  The blue short dashed line is the model MW-C
      (see Table 1) without threshold, with inside out formation. The
      model MW-D characterized by the absence of inside-out is
      indicated by the light-blue long dashed line.  The model MW-B
      with variable SFE, inside out formation and threshold is
      represented by the brown dashed line). The black dotted line
      represents the model MW-A with inside-out, threshold, and
      constant SFE. The data are from Luck \& Lambert (2011).}
		\label{altoredshiftORD}
\end{figure}

\section{The galactic habitable zone of the Milky Way and M31}
The galactic habitable zone (GHZ) is defined as the region with
sufficiently high metallicity  to form planetary systems in
which Earth-like planets could be found and might be capable of
sustaining life.
\begin{figure} 
	  \centering   
    \includegraphics[scale=0.4, angle=-90]{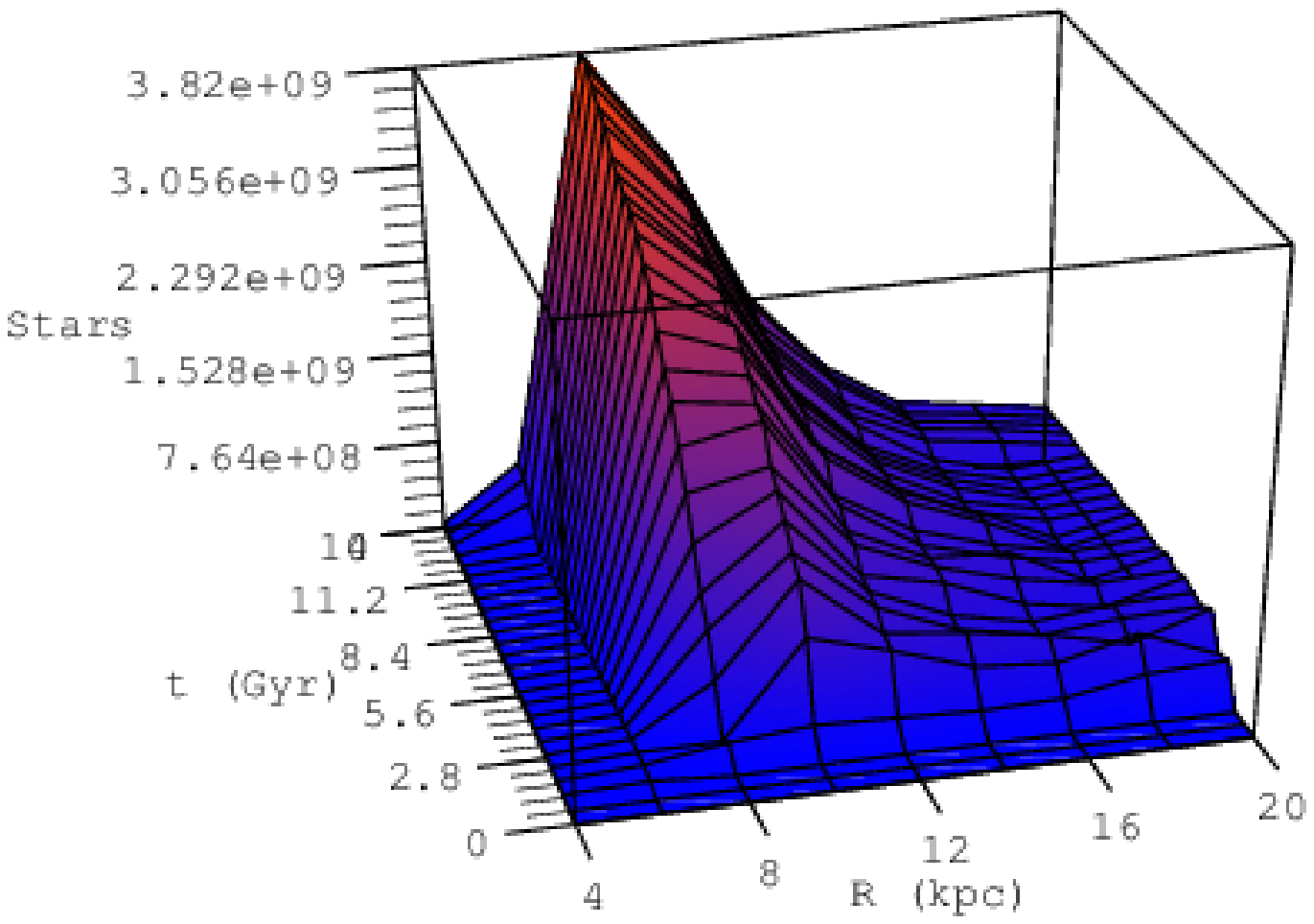} 
 \includegraphics[scale=0.4, angle=-90]{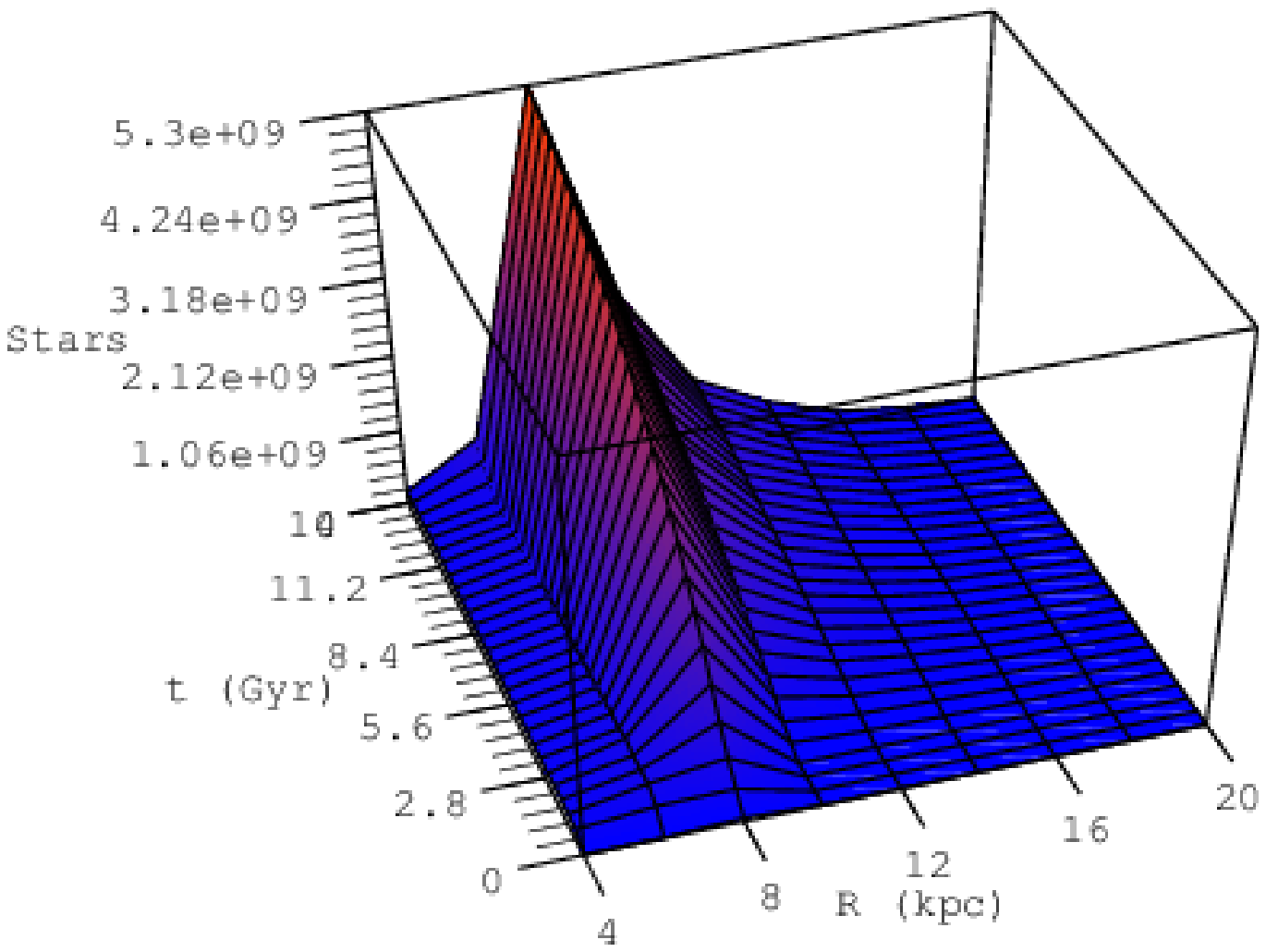} 
\centering \includegraphics[scale=0.4,angle=-90]{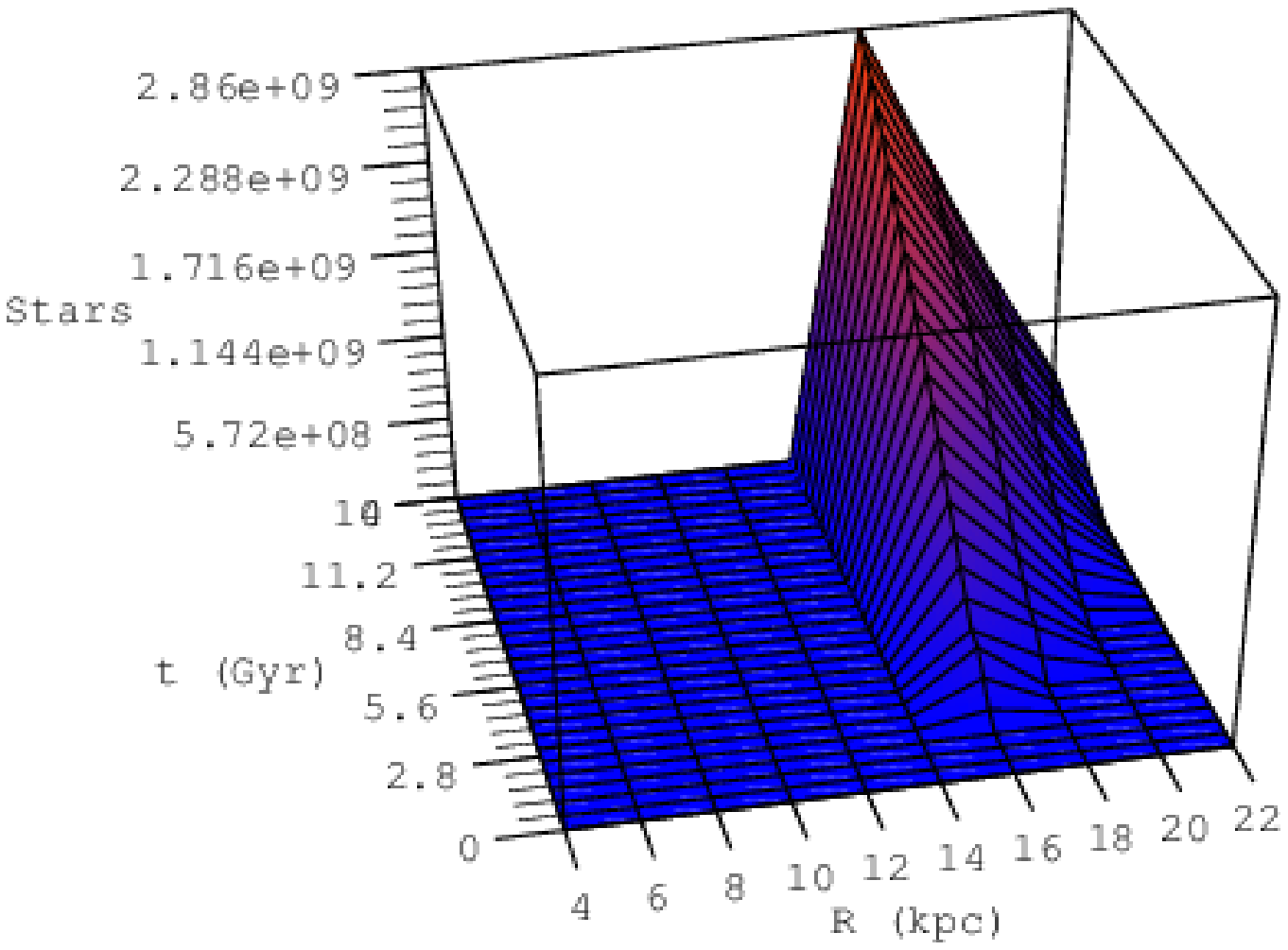}
  \includegraphics[scale=0.4, angle=-90]{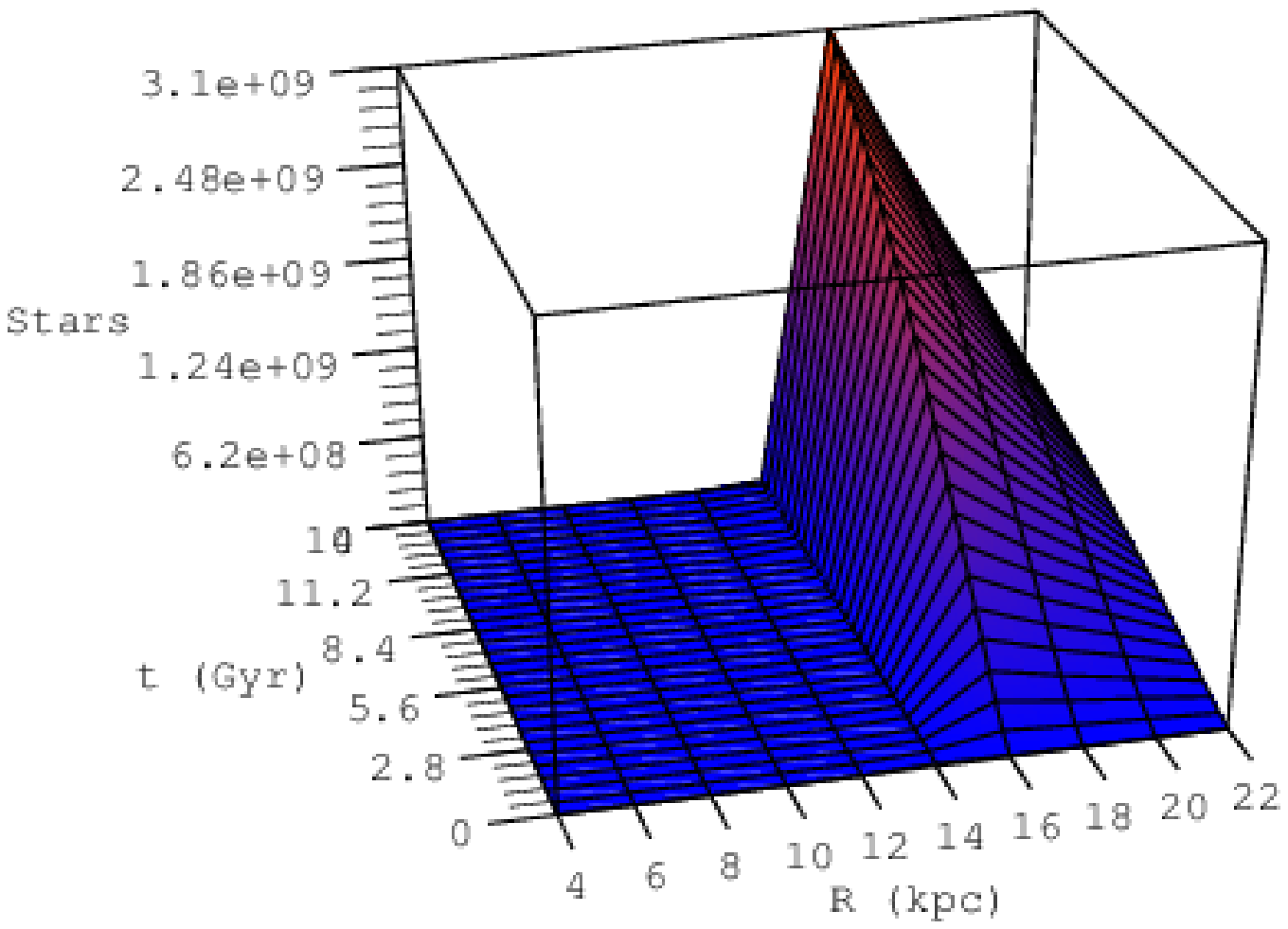} 
    \caption{The total number of stars having Earths ($N_{\star
        life}$) as a function of  galactocentric distance and 
      galactic time
      for the Milky Way ({\it Upper panels}) and for M31 ({\it Lower panels}). The ($N_{\star life}$) values are computed
      within concentric rings, 2 kpc wide. {\it Left panels:} 
      classical model results without radial gas flows are shown using the MW-A model for the Milky Way and the model M31-B for M31 (see Table 1). {\it
        Right panels:} model results with radial gas flows are
      reported adopting the MW-R model for the Milky Way and the model M31-R for M31 (see Table 1).} 
		\label{map_S2IT_SN2}
\end{figure} 
We have assumed that the probability of forming habitable Earth-like
planets depends on the [Fe/H] (following the prescriptions of Prantzos
2008), the SFR and the supernova rate of the studied region. We define
$P_{GHZ}(R,t)$ as the fraction of all stars having Earths (but no gas
giant planets) which survived supernova explosions as a function of
the galactic radius and time:

\begin{equation}
P(R,t)= \frac{\int_0^t SFR(R,t') P_E (R,t') P_{SN}(R,t') dt'}{\int_0^t SFR(R,t')dt'}.
\label{GHZ}
\end{equation}
This quantity should be interpreted as the relative probability to
have complex life around one star at a given position, as suggested by
Prantzos (2008).  In eq. (\ref{GHZ}), $P_{SN}(R,t')$ is the
probability of surviving to SN explosion, and $P_E (R,t')$ is the
probability of having stars with Earth-like planets but not gas giant
planets which destroy the Earth-like planets.  Finally, we define the
total number of stars formed at a certain time $t$ and galactocentric
distance $R$ hosting Earth-like planet with life as $N_{\star \, life}(R,t)=P_{GHZ}(R,t) \times N_{\star tot}(R,t)$,
where $N_{\star tot}(R,t)$ is the total number of stars created up to
time $t$ at the galactocentric distance $R$.  

In Fig. 4 we show the $N_{\star \, life}(R,t)$ values as a function of
the Galactocentric distance and the Galactic time for models without
radial gas flows for the Milky Way (MW-A) and M31 (M31-B). We compare
those results with the GHZ obtained including radial gas
flows: the MW-R model for the Milky Way and the model M31-R for M31
(see Table 1). 

For both the Milky Way and Andromeda, the main effect
of the gas radial inflows is to enhance the number of stars hosting a
habitable planet with respect to the ``classical'' model results, in
the region of maximum probability for this occurrence. This is due to
the increase of gas toward inner region because of radial inflows,
which leads to larger SFR values. We also recall that models with
radial gas inflows have no threshold in the star formation. All
results are obtained by taking into account the supernova destruction
processes. In particular, we find that in the Milky Way the maximum
number of stars hosting habitable planets is at 8 kpc from the
Galactic center, and the model with radial flows predicts a number
which is 38\% larger than what predicted by the classical model. For
Andromeda we find that the maximum number of stars with habitable
planets is at 16 kpc from the center and that in the case of radial
flows this number is larger by 10\% relative to the stars predicted by
the classical model.

\end{document}